\documentclass[aps,pre,twocolumn,english,superscriptaddress,showpacs,floatfix]{revtex4}

\usepackage{amsmath,bbm}
\usepackage{psfig,graphics,graphicx}
\usepackage{babel}

\begin{document}

%\preprint{}

\title{Topological constraints on
spiral wave dynamics in spherical geometries with inhomogeneous
excitability}

\author{J\"orn Davidsen}
\email[]{davidsen@mpipks-dresden.mpg.de}
\affiliation{Max-Planck-Institut f\"ur Physik Komplexer Systeme,
N\"othnitzer Strasse 38, 01187 Dresden, Germany}
\affiliation{Chemical Physics Theory Group, Department of
Chemistry, University of Toronto, Toronto, ON M5S 3H6, Canada}
\author{Leon Glass}
\email[]{glass@cnd.mcgill.ca}
\affiliation{Department of Physiology, McGill University,
Montreal, Quebec H3G 1Y6, Canada}
\author{Raymond Kapral}
\email[]{rkapral@chem.utoronto.ca} \affiliation{Max-Planck-Institut
f\"ur Physik Komplexer Systeme,
N\"othnitzer Strasse 38, 01187 Dresden, Germany}
\affiliation{Chemical Physics
Theory Group, Department of Chemistry, University of Toronto,
Toronto, ON M5S 3H6, Canada}

\date{\today}

\begin{abstract}
We analyze the way topological constraints and inhomogeneity in
the excitability influence the dynamics of spiral waves on spheres
and punctured spheres of excitable media. We generalize the
definition of an index such that it characterizes not only each
spiral but also each hole in punctured, oriented, compact,
two-dimensional differentiable manifolds and show that the sum of
the indices is conserved and zero. We also show that heterogeneity
and geometry are responsible for the formation of various spiral
wave attractors, in particular, pairs of spirals in which one
spiral acts as a source and a second as a sink --- the latter
similar to an antispiral. The results provide a basis for the
analysis of the propagation of waves in heterogeneous excitable
media in physical and biological systems.

\end{abstract}

%\keywords{}

\maketitle


\section{Introduction}
Geometry and inhomogeneity influence pattern formation in chemical
and biological systems. \cite{murray89,kapral} One example where
these two factors play a crucial role is in the experimental
observations of distinctive spiral wave dynamics on the surfaces
of spherical beads, which are excitable inhomogeneous chemical
media \cite{maselko89,maselko90}. A biological example is the
origin of abnormal cardiac rhythms in the human heart which depend
on the anatomical substrate. The heart possesses a complex
nonplanar geometry with multiple chambers, with holes
corresponding to valves and blood vessels. Some serious
arrhythmias are associated with circulating spiral waves similar
to those observed in chemical media \cite{winfree01}. Since an
abnormal anatomical substrate is a common finding in patients with
some types of cardiac arrhythmias, and interventions that modify
the anatomy are an accepted form of therapy \cite{stein02},
theoretical analyses of the interplay between geometry of the
substrate and dynamics may help in the therapy of cardiac
arrhythmias.

In this paper we study spiral wave dynamics on (punctured) spheres
with spatially inhomogeneous excitability. We show for punctured
spheres that the sum of indices which characterize each spiral has
to be zero. Moreover, we demonstrate that topological constraints
imposed by the spherical geometry and inhomogeneity in
excitability lead to the formation of pairs of spirals, with
distinctive transient dynamics or as stable attractors, in which
one spiral acts as a source and a second as a sink leading to a
source-sink pair under a broad range of conditions. Our results
explain the experimental observations of spirals on spherical
beads \cite{maselko89,maselko90}.  While we do not consider
detailed models of cardiac wave propagation, our results may apply
to some generic aspects of atrial arrhythmias because the thin
walls of the atria can be described as two-dimensional (2d)
inhomogeneous excitable media with specific geometrical features.
\cite{3d}

\section{An Index Theorem for Phase Singularities}

The mathematical description of spiral waves is based on the
notion of phase which in turn allows one to characterize spiral
waves by an index. From this description, a number of topological
results placing restrictions on spiral wave dynamics can be
derived. \cite{glass77,winfree83,sumners87,winfree01,cruz-white03}

With the exception of a finite number of singular points, with
each point in an orientable and compact two-dimensional
differentiable manifold $M$ we identify a unique phase lying on
the unit circle, $\Phi \in S^1$. The resulting phase map or phase
field is assumed to be continuously differentiable, except at the
singular points. The manifold can be triangulated \cite{hocking88}
(subdividing it into a set of polygons), where none of the edges
or vertices of the polygons pass through a singularity. The index,
$I$ (sometimes also called the topological charge or winding
number) of a curve $C$ bounding a polygon is found by computing
the line integral
\begin{equation}\label{top_charge}
    2 \pi I = \oint_C \nabla \Phi \cdot d{\bf l},
\end{equation}
where the polygon is always traversed in a clockwise orientation.
By continuity of $\nabla \Phi$, $I$ must be an integer. The index
of a singular point is uniquely defined as the index of any curve
$C$ provided that $C$ encircles the point but no other singular
points. The index of a curve that does not enclose any singular
points is obviously zero.

If the manifold $M$ has no boundaries, each edge of the
triangulation is an edge of two polygons. Since the line integral
adds up the change in phase along the various edges of the
polygon, the sum of the indices of the singular points for a phase
field in $M$ is
\begin{equation}\label{indextheorem}
\Sigma_M I =0,
\end{equation}
where the sum is over all the singular points.  This follows since
the contribution of the change in phase of each edge to the total
integral is counted twice, but since the edge is traversed in
opposite directions each time, the net contribution of each edge
is zero. This index theorem is applicable to tori, and other
surfaces of genus different from 0. However, unlike the more
familiar Poincar\'e index theorem (see Ref. \cite{guillemin74} p.
74 for vector fields) the sum of the indices of the singular
points does not depend on the genus of the surface.

This index theorem for manifolds without boundaries can be
extended to manifolds with boundaries. In the following, we will
consider the case of structures that arise from puncturing
orientable and compact two-dimensional differentiable manifolds.
The index of a hole can be uniquely defined as the index of a
curve $C$ provided that $C$ encircles the hole but no other holes
or singular points and $C$ is positively oriented with respect to
the domain which contains the hole and is bounded by $C$. Applying
this definition and taking the summation in Eq.
(\ref{indextheorem}) over the singularities and the hole, or the
holes if there is more than one hole, the index theorem can be
proven by the same line of arguments as for the case of manifolds
without boundaries.

This extension is important in the heart, for example, where the
atrium is punctured by valves and veins. In such cases one is led
to consider manifolds with holes; for example, a sphere with a
hole. A sphere with a hole is topologically equivalent to a disk,
and, indeed, the results for disks and for spheres with holes are
consistent: For the disk $D^2$, bounded by a curve $C$,
\begin{equation}\label{dindextheorem}
\Sigma_{D^2} I =\oint_C \nabla \Phi \cdot d{\bf l},
\end{equation}
so that the sum of the indices of the singular points in the disk
is equal to the index of the curve $C$ bounding the disk. If there
is a single singular point on the disk, with an index of +1, the
index of the curve bounding the disk will also be +1. Imagine now
the boundary of the disk to be brought together (like a
draw-string bag) so that the boundary of the disk now defines a
hole in the sphere. In this geometry the curve $C$ will be
traversed in an opposite orientation (the hole is now inside $C$)
from the direction it was traversed when it was the boundary of
the disk. Now if there is a singular point with an index of +1 on
the sphere, the index of the hole is -1, so that the sum of the
indices is again zero.

Since it is necessary to conserve the sum of the indices,
singularities of index $\pm 1$ usually arise and are destroyed in
pairs of opposite sign \cite{winfree83}. An exception occurs when
singularities are destroyed by collision into a boundary, so that
the index of the singular point and the index of the boundary both
change simultaneously. Another exception occurs if there are
singularities with index different from one. In such cases
interactions between different singularities can lead to
destruction or creation of odd numbers of singularities
\cite{zaritski02}.

\section{The FitzHugh-Nagumo Equation}

The FitzHugh-Nagumo (FHN) equation, \cite{numerics}
\begin{eqnarray}
\frac{\partial u}{\partial t} &=& \epsilon^{-1} \left( -
\frac{u^3}{3} + u - v \right) + D_u \nabla^2 u,\nonumber\\
\frac{\partial v}{\partial t} &=& \epsilon \left( u - \alpha v +
\beta \right) + D_v \nabla^2 v,\label{equdyn}
\end{eqnarray}
where $u({\bf r},t)$ and $v({\bf r},t)$ are two scalar fields,
$\epsilon^2$ is the ratio of the time scales associated with the
two fields, and $D_u$ and $D_v$ are the constant diffusion
coefficients, is a
prototypical model for excitable media.
We choose $D_u = 2$ and $D_v = 0$. The real
parameters $\alpha$ and $\beta$ characterize the local dynamics
and, hence, the local excitability. Whenever $0<\alpha<1$, $\alpha
\epsilon^2 < 1$ and $|\beta| > \beta_H \equiv (1 - \alpha
\epsilon^2)^{1/2} (1/3 (2 \alpha + \alpha^2 \epsilon^2) - 1)$, the
FHN system is excitable. At $\beta_H$ a Hopf bifurcation occurs
such that for $|\beta| < \beta_H$ the system exhibits
oscillations. In the following we take $\alpha = 0.2$, $\epsilon=0.2$ and
$\beta > \beta_H = 0.863\dots$.

We consider a spherical shell whose outer and inner radii are
$R_e$ and $R_i$, respectively, and focus on thin spherical shells
where $R_i=40$, $R_e=42$. The radii are large enough to avoid a
curvature-dependent loss of excitability \cite{davydov02}, and the
shell is sufficiently thin that the dynamics is effectively 2d
corresponding to the dynamics on a sphere. \cite{shell} The
initial condition is a domain of ``excited" state, adjacent to a
domain of the ``refractory" state. Both domains have the forms of
slices of the same size oriented from the north to the south pole
\cite{details} and yield a pair of counter-rotating spirals.

In order to apply the topological results to the FHN equation, it
is necessary to first define the phase. We define a phase,
$\Phi({\bf r},t)$, based on the equation $\tan{\Phi({\bf r},t)} =
v({\bf r},t)/u({\bf r},t)$ if $v({\bf r},t)\neq 0$ and $u({\bf
r},t)\neq 0$. Thus, singular points at given $t$ are points ${\bf
r}$ in the medium for which $v({\bf r},t)=0$ and $u({\bf r},t)=0$.
For each $t$, we obtain a continuously differentiable phase map
${\cal M}^t=\Phi({\bf :},t)|_{{\cal D}^t}$ that associates to each
point in a well-defined domain ${\cal D}^t$ a phase lying on the
unit circle, $\Phi \in S^1$. For our FHN medium, the domain is the
surface of a (punctured) sphere reduced by a finite number of
points where the phase is singular at fixed $t$.

Rotating spiral waves in the FHN equations are obviously
associated with a singular point which is called the spiral core.
In what follows, we assume that there are only single-armed
spirals so that a clockwise rotating source has an index of +1,
and a counterclockwise rotating source has index -1. A clockwise
rotating sink has an index of -1, and a counterclockwise rotating
sink has index +1. From Eq.~(\ref{indextheorem}), it is impossible
to have a single rotating spiral wave on a sphere; in addition,
there must be at least another singular point or a hole with
nonzero index.

For excitable media, a non-zero index of a hole implies that wave
fronts travel permanently around the hole such that the numbers of
fronts travelling clockwise and counterclockwise are different.
This includes the particularly simple case of a single wave front
travelling around the hole which can be considered as a spiral
wave associated with the hole.

\section{Spiral Wave Dynamics in Spherical Geometries}
\subsection{Dynamics in excitability gradients}
\begin{figure}
 \includegraphics[width=0.75\columnwidth]{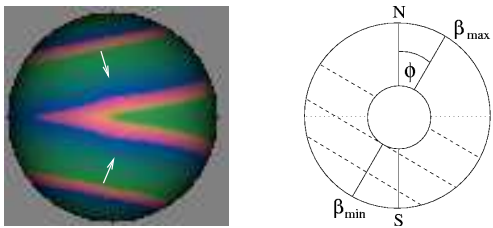}
 \caption{\label{exc_hom}
(Color online) Left: Spiral waves of excitation (light fronts) on
the sphere for constant $\beta=\beta_{ex}$ emanating from spiral
cores close to the poles on an equator projection. The white
arrows show the direction of propagation. One annihilation front
along the equator can be identified. Right: Sketch of the constant
gradient in the inhomogeneous case. The dashed lines are the
equi-$\beta$ lines and we choose $\beta_{max}=1.0$ and
$\beta_{min}=\beta_{ex}$. The angle $\phi$ describes the
orientation with respect to the axis from pole to pole. The
results described in the text do not depend qualitatively on the
choice of $\phi$ or $\beta_{min}$ and $\beta_{max}$ as long as
they yield stable spirals.}
 \end{figure}

First consider homogeneous FHN media with a constant
$\beta=\beta_{ex}\equiv0.9$. The wave fronts emanating from the
spiral cores with opposite index ``annihilate" along the equator
such that each spiral determines the dynamics on half of the
sphere (see Fig.~\ref{exc_hom}, left panel) --- similar to what
has been observed in Ref.~\cite{yagisita98} for a different
excitable system.
We have shown that this behavior is robust with respect to
disorder in excitability with small amplitude and correlation
length. If random, uncorrelated spatial variations in $\beta$
exist on length scales much smaller than the diameter of the
spiral core meander \cite{random_details}, the dynamics is able to
average over such small-scale inhomogeneities. The robustness
explains why such states have been experimentally observed in some
chemical reactions on spherical beads which are intrinsically
inhomogeneous \cite{maselko90}.

\begin{figure}
 \includegraphics[width=\columnwidth]{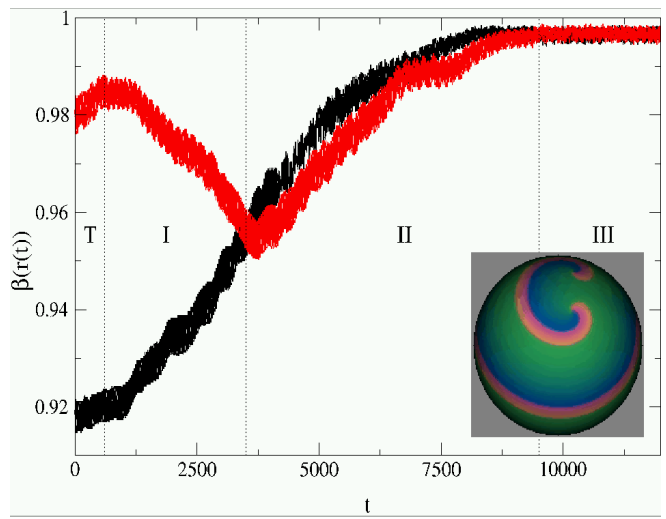}
 \caption{\label{beta}
(Color online) The local excitability $\beta({\bf r}(t))$ at the
spiral cores versus time. Gradient-induced motion of the two
spiral cores leads to a change in the local excitability at the
cores with time. The spiral period in the final state is
$T_0=13.2\pm0.1$. Four different regimes can be identified (see
text). Inset: The final bound pair of counter-rotating spirals in
regime III for $\phi=51.0^\circ$ is shown on an equator projection
such that the point of lowest excitability lies on the central
longitude. The spiral closer to the equator has index $-1$ while
the other one has index $+1$.}
 \end{figure}

\begin{figure}
 \includegraphics[width=\columnwidth]{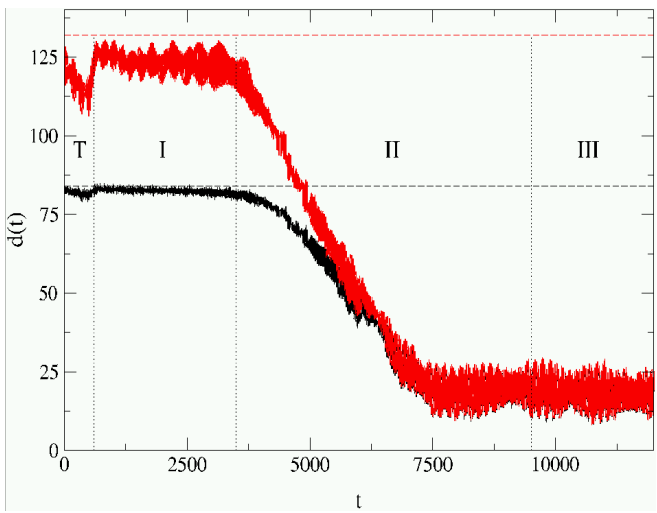}
 \caption{\label{d}
(Color online) The distance between the two spiral cores $d(t)$
versus $t$. Gradient-induced motion of the two spiral cores leads
to a change in the distance $d$ between the cores with time. The
lower and upper curves correspond to the distance in
$\mathbbm{R}^3$ and $S^2$, respectively. The dashed lines are the
respective upper bounds given by the size of the sphere.}
 \end{figure}

\begin{figure}
 \includegraphics[width=\columnwidth]{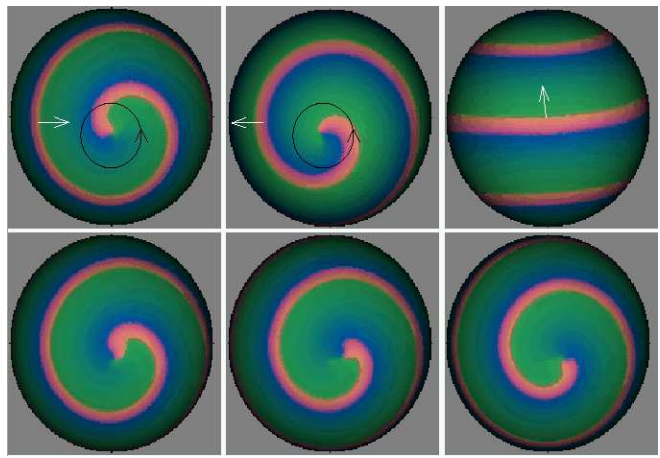}
 \caption{\label{anti_series}
(Color online) Waves of excitation on the sphere in regime I of
the gradient-induced dynamics shown in Figs.~\ref{beta} and
\ref{d}. A source-sink pair has formed. For random spatial
variations of excitability with a correlation length comparable to
the diameter of the spiral core meander, the final state is very
similar to the one shown here \cite{random_details}. Upper panel
from left to right: View centered at the north pole, south pole
and the equator. The source at the south pole has index $-1$ and
the sink at the north pole index $+1$. The black circles show
possible choices of $C$. The white arrow shows the direction of
wave propagation. Lower panel: Dynamics at the north pole. Time
increases from left to right with $\Delta t = 2.5$ between
snapshots.}
 \end{figure}

Applying a constant gradient in $\beta$ as sketched in the right
panel of Fig.~\ref{exc_hom} leads to a different scenario. The time
evolution of the spiral pair may be partitioned into four distinct
regimes as shown in Figs.~\ref{beta} and \ref{d}.
Because of the gradient, the frequencies of the two spirals differ
since a higher value of $\beta$ corresponds to lower excitability,
which generally implies a lower spiral frequency \cite{winfree91}.
During a short transient T, the spiral with the higher frequency
assumes control of the dynamics \cite{krinsky83} on the sphere.
The location on the sphere, where wave fronts emanating from the
two spiral cores annihilate, moves toward the core of the
low-frequency spiral. Finally, the low-frequency spiral core is
pushed farther from the high-frequency spiral core
\cite{krinsky83,ermakova86}(see Fig.~\ref{d}). After this short
transient, the wave fronts travel from pole to pole leading to the
creation of a source-sink pair. This (intermediate) state is shown
in Fig.~\ref{anti_series} and corresponds to regime I in
Figs.~\ref{beta} and \ref{d}. Viewed from the low excitability end
of the sphere, the waves wind into a small region about the core,
reminiscent of the structure of antispirals, i.e., inward moving
spirals seen in oscillatory media \cite{gong03}. However, the
origin of this inward spiral motion in oscillatory media differs
and is distinct from that observed here. In oscillatory media,
either spirals or antispirals are stable depending on system
parameters and the wave length diverges on the border in the
parameter space between these two regimes. Thus, antispirals exist
independently of spirals. This is not the case here because the
generation of an inward moving spiral relies on the presence of a
spiral source and spherical geometry. For example, consider the
FHN system with a disc geometry and a radial gradient in
excitability such that the maximum value of $\beta$ is located in
the center of the disc. In this case, a source-sink pair cannot
occur because the high-frequency spiral, acting as a source, would
push the other (low-frequency) spiral out of the system, excluding
the presence of any strong random inhomogeneities in excitability
which may pin the low-frequency spiral and prevent its motion
(see, e.g., \cite{biktashev01}). The lower panel of
Fig.~\ref{anti_series} shows that remnants of the low frequency
spiral persist in a small area around the core. It has been
speculated that antispiral waves might occur in cardiac tissue
\cite{vanag01}. While excitable media cannot support a regime of
exclusive antispirals due to their excitable character, our
results show that source-sink pairs with similar characteristics
could form in the heart where the underlying dynamics is
excitable, the medium is inhomogeneous and the topology is similar
to a (punctured) sphere.

Spiral dynamics of the type described above has been observed by
Maselko and Showalter \cite{maselko89} in experiments on the
excitable Belouzov-Zhabotinsky reaction on spherical beads. They
attributed the generation of spiral source-sink pairs to
inhomogeneities in the medium related to differing chemical
environments. This is consistent with our findings for systems
with gradients and is further confirmed by the work reported in
Ref.~\cite{yagisita98}. While the generation of source-sink pairs
due to a gradient in the FHN medium investigated here is only an
intermediate state (but one that persists for approximately 240
spiral periods in our simulations), random spatial variations of
the excitability with a correlation length comparable to the
diameter of the spiral core meander or larger can lead to a final
state consisting of such a source-sink pair \cite{random_details}.
This is due to the fact that the source can be trapped in a region
of depressed local excitability.

Not only does the gradient in the FHN medium change the local
excitability but it also induces a drift of the spiral cores
\cite{biktashev95}. For our model, the drift is rather slow
compared to the transition to the source-sink pair which takes
place during the transient regime T. This can be seen in
Fig.~\ref{beta}. (The fluctuations in $\beta({\bf r}(t))$ are due
to spiral meandering.) In regime I, the dominating spiral drifts
toward lower excitability and its wave fronts continuously push
the other core in the opposite direction, thus, keeping the
distance $d$ between the cores approximately constant (see
Fig.~\ref{d}). The fluctuations in $d(t)$ are again due to
meandering of the spiral cores. Because of this drift, the local
excitabilities at the two spiral cores approach each other until
they become equal.

At this point, regime II in Figs.~\ref{beta} and \ref{d} is
entered. The dynamics change drastically: the slaved spiral
reverses its drift direction and regains control over its own
dynamics. Both spirals drift toward lower excitability. Due to the
geometric constraints imposed by the spherical geometry, the
spirals approach each other until they form a bound pair (at $t
\approx 7500$).

They finally reach a stable state (at $t \approx 9500$)
corresponding to regime III in Figs.~\ref{beta} and \ref{d}.
Neither the \emph{average} distance between the spirals nor the
\emph{average} local excitability changes further. Yet, on top of
the persisting unsynchronized meandering of the two spiral cores,
the bound pair slowly moves along a (closed) equi-$\beta$ curve on
the surface of the sphere. The direction of the motion depends on
the initial condition, i.e., whether the spiral with positive
index was initially closer to the point of lowest excitability or
to the point of highest excitability than the spiral with negative
index. The wave dynamics generated by this bound pair is shown in
the inset of Fig.~\ref{beta}.

While kinematic theory applies only to spirals with large cores,
it is instructive to note that this theory predicts that the
direction of the drift due to gradients depends on the model
system and its parameters \cite{mikhailov94}. Although our
simulations have shown the same direction of the drift for a range
of parameters in the meander region of the FHN phase diagram, it
is conceivable that, under different circumstances favoring a
drift toward higher excitability, one spiral could act as a
permanent source and a source-sink pair could be the final
attractor. Such a scenario would also be consistent with the
experimental findings in Ref. \cite{maselko89}.

\subsection{Punctured spheres}
\begin{figure}
  \includegraphics[width=\columnwidth]{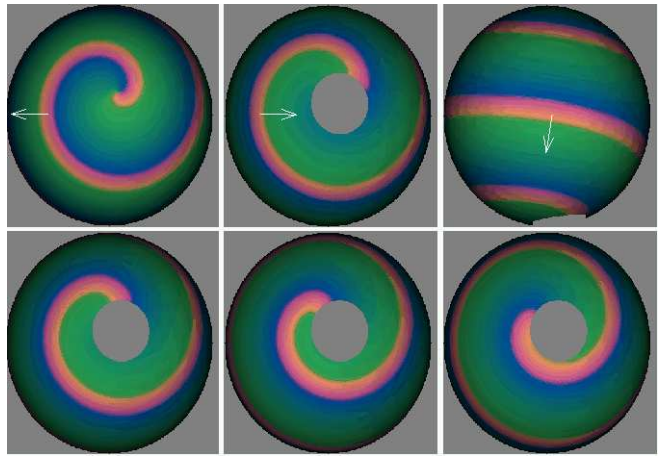}
  \caption{\label{series_hole}
(Color online) Waves of excitation on the punctured sphere
propagating towards the hole. A source-sink spiral wave pair has
formed in the final state. Upper panel from left to right: View
centered at the north pole, south pole and the equator. Lower
panel: Dynamics at the south pole. Time increases from left to
right with $\Delta t = 2.5$ between snapshots.}
\end{figure}

Next, we consider a homogeneously excitable sphere with a single
hole. Two scenarios can be observed depending on the location of
the hole with respect to the spiral pair. If a spiral wave is not
permanently attached to the hole, the dynamics is very similar to
the case without any hole. If one of the spirals is permanently
attached to the hole, the frequency of this spiral is lowered. The
size of the hole determines the frequency of the spiral because
the wave front has to travel around the hole. The transient
dynamics is similar to that in regime T for the case with a
gradient; however, no drift of the spiral cores is induced and the
final state is a spiral source-sink pair as shown in
Fig.~\ref{series_hole}. Not only is the net index conserved during
the transition to a spiral source-sink pair but so is the index of
the individual spirals: during the transition an outgoing
counterclockwise (clockwise) oriented spiral is converted into an
ingoing clockwise (counterclockwise) oriented spiral. Thus, the
formation of spiral source-sink pairs conforms with the
topological constraints.

If a gradient as well as a hole is present, the spiral drift
discussed earlier also determines the final state, which depends
on the hole's location in the gradient field. For instance, if the
point of lowest excitability in the medium is on the hole
boundary, simulations show that the spiral which is not attached
to the hole will stabilize close to this point. In this final
state odd numbers of wavefronts are attached to the hole, again
conserving the topological charge of the hole.

\section{Concluding Remarks}
Inhomogeneities due to spatially-varying excitability on
(punctured) spherical shells lead to complex spiral wave dynamics
and the formation of source-sink spiral pairs in excitable media.
The results presented here are immediately applicable to excitable
media in more complicated geometries such as tori or multi-holed
tori and to situations in which multi-armed spirals are found.
This includes mathematical modelling of cardiac tissue. The
approach taken in this paper stresses constraints and aspects that
apply to, and must be observed in, all realistic models of the
heart satisfying certain criteria of continuity. There are also
implications for the treatment of cardiac arrhythmias. In
cardiology it is sometimes possible to develop maps showing the
timing of the excitation over limited regions of heart
\cite{stein02}. In this case, a sink might be confused for a
source (of the arrhythmia), and this might have implications for
the diagnosis of the mechanism and the choice of therapy. The
current work shows how partial knowledge about what is happening
in some regions that could be observed, might be helpful in
establishing properties of dynamics that could not be observed.
While the types of sinks we have described here have only been
observed in chemical media \cite{maselko89,maselko90} so far, we
certainly expect their existence in the cardiological domain.

We thank F. Chavez, M. B\"ar, S. G. Whittington and D. Sumners for
helpful discussions and G. Rousseau for providing numerical tools.
This work was supported in part by a grant from MITACS.

%\appendix
%\section{}

%\begin{acknowledgments}
%\end{acknowledgments}



\begin{thebibliography}{26}
\bibitem[{num()}]{3d}
\bibinfo{note}{Some aspects of cardiac arrhythmias may arise from
scroll wave filament instabilities which depend on the 3d nature
of cardiac tissue \cite{winfree87,mikhailov03}.}
\bibitem[{num()}]{numerics}
\bibinfo{note}{We solve Eq. (\ref{equdyn}) numerically using an algorithm that
  automatically adjusts the time step to achieve an effcient simulation while
  controling the error in the solution \cite{rousseau00}.}
\bibitem[{det()}]{shell}
\bibinfo{note}{It is computationally convenient to carry out the calculations
in a thin spherical shell constructed from a cubic rectangular
grid. The shell thickness is small compared to characteristic
wavelengths and diffusion lengths so that the small shell
thickness does not influence the spiral wave dynamics.}
\bibitem[{det()}]{details}
\bibinfo{note}{The initial conditions are as in Ref.~\cite{chavez01} with
  spherical coordinates $\Delta\theta=\pi/2$ and $\Delta\phi=\pi$.}
\bibitem[{ran()}]{random_details}
\bibinfo{note}{For a Gaussian distribution of $\beta$ in boxes of volume
  $0.5^3$ with mean 0.95 and standard deviation 0.013, the diameter of the
  spiral core meander is roughly 10 space units.}
\end{thebibliography}
\end{document}